\documentclass[12pt]{iopart}
\usepackage{braket,color,graphicx,multirow,subfig,url}
\newenvironment{bmatrix}{\left[\begin{array}{*{20}{c}}}{\end{array}\right]}

\newcommand{\pa}[2]{\partial #1/\partial #2}
\newcommand{\Paa}[2]{\frac{\partial^2 #1}{\partial #2^2}}
\newcommand{\Pab}[3]{\frac{\partial^2 #1}{\partial #2\, \partial #3}}
\begin{document}
\title{Strategic Insights From Playing the Quantum Tic-Tac-Toe}
\author{J. N. Leaw and S. A. Cheong}
\address{Division of Physics and Applied Physics,
School of Physical and Mathematical Sciences,
Nanyang Technological University,
21 Nanyang Link, Singapore 637371}
\ead{cheongsa@ntu.edu.sg}

\begin{abstract}
In this paper, we perform a minimalistic quantization of the classical game of
tic-tac-toe, by allowing superpositions of classical moves.  In order for the
quantum game to reduce properly to the classical game, we require legal quantum
moves to be orthogonal to all previous moves.  We also admit interference
effects, by squaring the sum of amplitudes over all moves by a player to compute
his or her occupation level of a given site.  A player wins when the sums of
occupations along any of the eight straight lines we can draw in the $3 \times
3$ grid is greater than three.  We play the quantum tic-tac-toe first randomly,
and then deterministically, to explore the impact different opening moves, end
games, and different combinations of offensive and defensive strategies have on
the outcome of the game.  In contrast to the classical tic-tac-toe, the
deterministic quantum game does not always end in a draw.  In contrast also to
most classical two-player games of no chance, it is possible for Player 2 to
win.  More interestingly, we find that Player 1 enjoys an overwhelming quantum
advantage when he opens with a quantum move, but loses this advantage when he
opens with a classical move.  We also find the quantum blocking move, which
consists of a weighted superposition of moves that the opponent could use to win
the game, to be very effective in denying the opponent his or her victory.  We
then speculate what implications these results might have on quantum information
transfer and portfolio optimization.

\end{abstract}


\maketitle

\section{Introduction}
\label{sect:intro}

Since Bouwmeester \emph{et al.}'s 1997 empirical demonstration 
of quantum teleportation \cite{Bouwmeester1997Nature390p575}, first
proposed theoretically by Bennett \emph{et al.}
\cite{Bennett1993PhysRevLett70p1895}, there has been a surge of
interest in quantum information transfer between two parties, Alice
and Bob (see for example, \cite{Nielsen1998Nature396p52,
Furusawa1998Science282p706, Pan2001Nature410p1067,
Riebe2004Nature429p734, Barrett2004Nature429p737,
Chaneliere2005Nature438p833}, and the reviews
\cite{Bennett2000Nature404p247, Galindo2002RevModPhys74p347,
Braunstein2005RevModPhys77p513}).  At the same time, quantum
cryptography research has been focussed on devising ways to prevent a
third party, Eve, from intercepting and reading the message
transmitted over a quantum channel, or for Alice or Bob to detect any
attempt at eavesdropping \cite{Bennett1984IntConfCompSysSigProc,
Ekert1991PhysRevLett67p661, Bennett1992PhysRevLett68p557,
Bennett1992PhysRevLett68p3121} (see review by Gisin \emph{et al.}
\cite{Gisin2002RevModPhys74p145}).  But what if Eve, frustrated at
failing in every attempt to decipher Alice's message to Bob, turn her
attention to foiling all transmissions?  Should this quantum jamming
scenario develop, Alice will be forced to explore various strategies
to get her message through to Bob, knowing that Eve will attempt to
interrupt the transmission, but not knowing beforehand how she plan to
do so.

In essence, cutting the measurements Bob has to make out of the
picture, the ding-dong decisions made by Alice and Eve have the
flavour of a two-player game.  Naturally, because information is
transferred across quantum channels, this is a quantum game, not a
classical game.  Adding quantum-mechanical elements to a classical
game always lead to surprises.  In 1999, Meyer constructed a quantum
game of penny flip, and concluded that quantum strategies increase a
player's payoff beyond what is possible with classical strategies
\cite{Meyer1999PhysRevLett82p1052}.  Eisert \emph{et al} later
analyzed non-zero-sum games and found for the famous Prisoner's
Dilemma that the the classical dilemma no longer arise if quantum
strategies are allowed \cite{Eisert1999PhysRevLett83p3077}.  Since
these pioneering works, there have been further studies on the exact
nature of quantum advantages \cite{Du2002ChinPhysLett19p1221,
Zhao2004ChinPhysLett21p1421, Aharon2008PhysRevA77e052310}, whether
these advantages persist when the games are noisy
\cite{Johnson2001PhysRevA63e020302R, Chen2002PhysRevA65e052320,
Guinea2003JPhysAMathGen36pL197, Flitney2005JPhysAMathGen38p449}, and
how entanglement influences the choice of quantum strategies
\cite{Du2005JPhysAMathGen38p1559, Yukalov2010EurPhysJB71p533,
Yukalov2010TheoryDecision}.  These works also spawned a series of
in-depth studies into the game-theoretic structure of quantum games
\cite{Iqbal2002PhysRevA65e022306, Lee2003PhysRevA67e022311,
Nawaz2004JPhysAMathGen37p11457, Arfi2005TheoryDecision59p127,
Ozdemir2007NewJPhys9p43, Ichikawa2008JPhysAMathTheor41p135303}.

The quantum information transfer scenario described above is an asymmetric
two-player quantum game, because the moves available to Alice are not the same
as those available to Eve.  In the financial arena, portfolio optimization can
also be viewed as a symmetric $N$-player quantum game, in the sense that the
same set of moves are available to all $N$ players.  Here, stocks are the
classical states, and portfolios made up of linear combinations of long and
short positions on these stocks are the quantum states.  When one fund manager
optimizes his portfolio, the optimalities of all other portfolios are affected,
forcing the other fund managers to also adjust their portfolios.  In this sense,
the stock market is a gigantic real-time multiplayer game where a large number
of fund managers reacts to price changes induced by other fund managers, making
adjustments to keep their portfolios optimal.  This is an area where the
relatively young field of quantum game theory can potentially make important
contributions.

To understand at a deeper level how quantum mechanics influence the choice of
strategies for such games, and eventually their outcomes, we analyze the
simplest two-player game of tic-tac-toe.  In Section \ref{sect:quantum}, we will
define the quantum moves and winning condition that we have adopted, and explain
how these are similar to or different from existing quantizations of the game.
In Section \ref{sect:randomgames}, both players make random moves allowed by our
rules, to simulate a benchmark situation where there is total absence of
strategy, for comparison against the random classical game.  We find that Player
1 wins about 60\% of the time in both random games, but Player 2 is at a greater
disadvantage in the quantum game.  We then study the impacts of different
opening moves on the random games, to find classical opening moves being most
favourable towards Player 2.  We also study end-game situations, where Player 1
is on the verge of winning, i.e.~Player 1 will surely win on the next move, if
Player 2 forfeits his or her move.  Here we find that Player 2 can effectively
deny Player 1 of his victory, by playing a blocking move comprising a weighted
superposition of the best moves that Player 1 can make to win.  Based on our
understanding derived from the random games, we then analyze in Section
\ref{sect:deterministicgames} the effectiveness of different strategies that the
two players can adopt in deterministic games.  For all strategy pairs, the
outcomes are very similar: Player 2 wins more deterministic games than Player 1,
when Player 1 opens with a classical opening move.  On the other hand, when
quantum opening moves are used, the natural advantage to Player 1 is restored,
with Player 2 winning only a small, but non-zero, proportion of deterministic
games.  Finally, we summarize our most important findings in the Section
\ref{sect:conclusions}.

\section{Quantum moves and winning condition}
\label{sect:quantum}

The classical tic-tac-toe is a childhood game played on a $3 \times 3$
grid.  It is a two-player game of no chance, as no randomizing devices
(for example, a dice) are used.  In addition, it is also a game with
no hidden information (unlike, for example, the hands of opponents in
most card games).  Both players know what moves have been played, and
what moves are available to themselves, as well as to their opponents.
In this game, the two players take turn occupying empty sites on the
$3 \times 3$ grid.  A player wins whenever he succeeds in occupying a
straight line consisting of three sites, be it horizontally,
vertically, or diagonally.  Alternatively, if all nine sites are
occupied and no player succeeded in making a line of three sites, then
the game ends in a draw (also called a tie).  In fact, if both players
make no mistakes, it can be proven mathematically that the classical
tic-tac-toe always ends in a draw \cite{Berlekamp2003WinningWays}.

To quantize games for two or more players, generalized quantization
schemes have been proposed \cite{Nawaz2004JPhysAMathGen37p11457,
Ozdemir2007NewJPhys9p43}.  These game-theoretic quantization schemes
allow us to very quickly construct payoff matrices, but they are not
convenient for implementing iterated play where the space of moves
diminishes with every move made.  The quantization scheme we chose is
very similar to that defined by Goff \emph{et al.}
\cite{Goff2002Proc38thAIAAConf, Goff2006AmJPhys74p962}, but differs in
important aspects of iterated play.  Goff \emph{et al.}~developed
their version of the quantum tic-tac-toe as a teaching metaphor for
entanglement and measurement in quantum mechanics, and thus their main
interest is in introducing measurement, and the ensuing wave function
collapse, into the game.  However, when we play by Goff \emph{et
al.}'s rules, the quantum tic-tac-toe does not properly reduce to the
classical game upon the restriction to classical moves.  In the
subsections to follow, we will introduce a set of rules that embodies
part of the essence of being `quantum', but at the same time properly
reduces to the classical rules when only classical moves are used.

\subsection{The quantum move} 

\begin{figure}[htbp]
\centering
\includegraphics[scale=0.5]{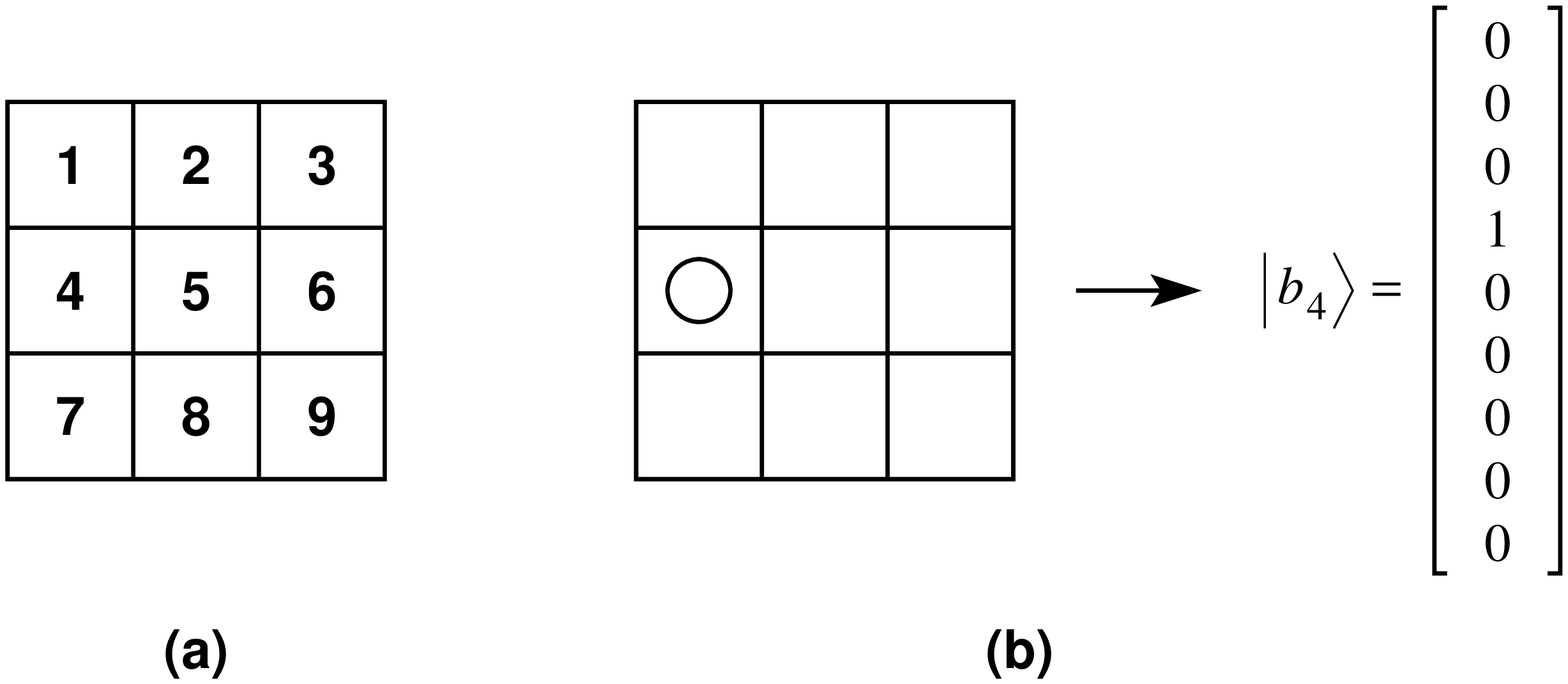}
\caption{The (a) sites on the $3 \times 3$ grid for tic-tac-toe,
numbered from 1 through 9, and (b) an example of how a classical move
is mapped to a basis vector in the nine-dimensional vector space.}
\label{fig:basis}
\end{figure}

As with Goff \emph{et al.}, we map the nine possible classical moves
to basis vectors in a nine-dimensional vector space, as shown in
Figure \ref{fig:basis}.  However, in contrast to Goff \emph{et al.},
whose quantum moves partially occupy only two sites, we define our
\emph{quantum move}
\begin{equation}
\ket{m} = \sum_{i=1}^9 v_i \ket{b_i}, \quad
\sum_{i=1}^9 |v_i|^2 = 1
\end{equation}
to be any normalized linear combination of the classical moves
$\{\ket{b_i}\}$, i.e.~we allow simultaneous partial occupation of any
number of sites.  In general, the amplitudes $v_i$ can be complex.  In
this paper, we restrict ourselves to real $v_i$, to make the
numerical studies presented in Sections \ref{sect:randomgames} and
\ref{sect:deterministicgames} simpler.

For our quantum tic-tac-toe to properly reduce to the classical
tic-tac-toe, we must impose the following restriction onto our quantum
moves.  In the classical game, a player may not play the classical
move $\ket{b_i}$, if it has already been played earlier.  This would
correspond to him or her trying to occupy an already occupied site.
Instead, he or she must play a classical move $\ket{b_j}$, with $j
\neq i$, if it has not been played.  Noting that $\ket{b_j}$ is by
construction orthogonal to $\ket{b_i}$, we require a legal quantum
move to be orthogonal to all previous quantum moves.  If we use
$\ket{m_{k\sigma}}$ to denote the $k$th quantum move made by player
$\sigma$, then the orthogonality requirement can be written as
\begin{equation}
\braket{m_{l\sigma}|m_{k\sigma}} = 0, \quad
\braket{m_{l'\sigma'}|m_{k\sigma}} = 0,
\end{equation}
for $l, l' < k$ and $\sigma' \neq \sigma$.  Here $\sigma = 1, 2$, and
$1 \leq k \leq 5$ for Player 1 and $1 \leq k \leq 4$ for Player 2.

\subsection{The winning condition}  

In Goff \emph{et al.}'s version of the quantum tic-tac-toe, the two players take
turns playing quantum moves of the form $\ket{m} = \frac{1}{\sqrt 2}\ket{b_i} +
\frac{1}{\sqrt 2}\ket{b_j}$, where $i \neq j$, until a closed loop of moves have
been made by one of the players.  The other player must then perform a
measurement on one site within the closed loop of moves, to collapse the state
of the game onto a classical state.  The classical state is then checked against
the classical winning condition, to see if one or the other player wins.  Else
the game continues, with the restriction that future quantum moves cannot occupy
any site on the collapsed loop.  The outcome of the game depends on which site
on the closed loop the wave function collapse started, and is thus not
deterministic.  For the quantum information transfer and portfolio optimization
scenarios outlined in Section \ref{sect:intro}, we prefer to have no wave
function collapse.  More importantly, we would like to define a deterministic
winning condition that is compatible with the quantum moves defined in the
previous subsection, and will also properly reduce to the classical winning
condition.  At the same time, we want to admit the possibility of
quantum-mechanical interference in our quantum game.

To define the winning condition, let us first define the \emph{weight}
$W_{pqr}^{k\sigma}$ Player $\sigma$ has along the straight line
through sites $p$, $q$, and $r$ after $k$ quantum moves.  In spite of
the orthogonality constraint described earlier, he or she is likely to
have played nonzero amplitudes at all sites for all $k$ moves.  To
compute the different occupation levels of the nine sites, we sum all
$k$ moves of Player $\sigma$,
\begin{eqnarray}
\fl
\ket{m_{1\sigma}} + \ket{m_{2\sigma}} + \cdots + \ket{m_{k\sigma}} &=
\sum_{i=1}^9 v_{i1\sigma} \ket{b_i} + \sum_{i=1}^9 v_{i2\sigma}
\ket{b_i} + \cdots + \sum_{i=1}^9 v_{ik\sigma} \ket{b_i} \\
&= \sum_{i=1}^9 \left(v_{i1\sigma} + v_{i2\sigma} + \cdots +
v_{ik\sigma} \right) | b_i \rangle \\
&= \sum_{i=1}^9 \left(\sum_{l=1}^k v_{il\sigma} \right) \ket{b_i},
\end{eqnarray}
where $v_{il\sigma}$ denotes the amplitude contribution to site $i$ by
the $l$th quantum move.  The term in the parentheses is the
accumulated amplitude in site $i$.  The weight $W_{pqr}^{k\sigma}$
Player $\sigma$ has along the \emph{direction} $pqr$ can then be
calculated as
\begin{equation}
\label{eq: weight}
W_{pqr}^{k\sigma} = \sum_{i=p,q,r} \left(\sum_{l=1}^k v_{il\sigma}
\right)^2.
\end{equation}
Thus, Player $\sigma$ wins after his or her $k$th move, if
\begin{equation}\label{eqn:quantumwinningcondition}
W_{pqr}^{k\sigma} \geq 3
\end{equation}
for some direction $pqr$.  For the sake of clarity in the rest of the
paper, we will refer to Player 1 in the masculine, and to Player 2 in
the feminine.

\section{Random games}
\label{sect:randomgames}

Even though our quantum tic-tac-toe `contains' the classical
tic-tac-toe, it is a very different game from its classical
counterpart.  In fact, it is so different we did not know how to play
it at first.  When two players play the game without any proper
strategy, the game would look very much like a random game.
Therefore, to start understanding our quantum tic-tac-toe, we played
random classical and quantum games, to see how different they really
are from each other.  This will also serve as a benchmark study of the
quantum game played in the absence of any strategy, for later
comparison against the deterministic strategic plays studied in
Section \ref{sect:deterministicgames}.

In a random classical game, the nine classical moves
$\{\ket{b_i}\}_{i=1}^9$ are played in random order.  After each move,
the maximum weight 
\begin{equation} 
W_{\max} = \max_{pqr} W_{pqr}
\end{equation} 
of the active player is calculated.  If this weight is equal to three,
the active player wins.  Otherwise, the game continues, until one
player wins, or the game ends in a draw.  In a random quantum game, we
first construct nine random vectors which are neither normalized nor
orthogonal.  We then apply the Gram-Schmidt orthonormalization
procedure on the nine vectors to obtain a set of nine orthonormal
\emph{random (quantum) moves}.  These random moves are then played
sequentially, until one player wins according to the quantum winning
condition in Eqn.~(\ref{eqn:quantumwinningcondition}), or the game
ends in a draw.

\subsection{Winning proportions}

After playing 10,000 random classical games and 10,000 random quantum games, we
tabulate the outcomes in Table \ref{tab: table1}.  In both the random classical
and random quantum games, Player 1 wins about 60\% of the time.  However, Player
2 is at a greater disadvantage in the random quantum game, in the sense that she
wins only 14.2\% of the time, as opposed to 28.5\% of the time in the random
classical game.  Furthermore, we see that in the random classical game, both
Player 1 and Player 2 win about 9\% of the time after their third move.  In the
random quantum game, no player wins after the third move.

\begin{table}[ht]
\caption{Outcomes of 10,000 random classical games and 10,000 random
quantum games.  Here we show the proportions of wins by Player 1 and
Player 2 after move $k$ for both games.  Player 2 has only four moves,
so the number shown for $k = 5$ is the proportion of games ending in a
draw.}
\label{tab: table1}
\begin{center}
\begin{tabular}{|c||c|c||c|c|} \hline
\multirow{2}{*}{Move $k$} & \multicolumn{2}{c||}{Claissical Game (\%)} & \multicolumn{2}{|c|}{Quantum Game (\%)} \\ \cline{2-5}
& Player 1 & Player 2 & Player 1 & Player 2 \\ \hline \hline
1 & 0 & 0 & 0 & 0 \\ \hline
2 & 0 & 0 & 0 & 0 \\ \hline
3 & 9.4 & 9.0 & 0 & 0 \\ \hline
4 & 26.5 & 19.5 & 21.8 & 14.2 \\ \hline
5/draw & 22.4 & 13.2 & 38.5 & 25.5 \\ \hline
\end{tabular}
\end{center}
\end{table}

\begin{figure}[ht]
\begin{center}
\includegraphics[scale=0.5]{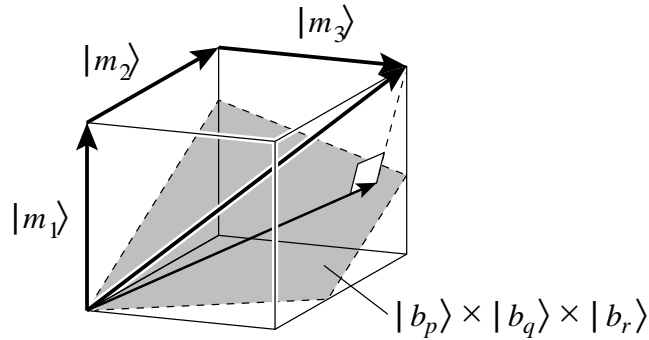}
\end{center}
\caption{Schematic diagram showing the resultant vector for three orthonormal
quantum moves $\ket{m_1}$, $\ket{m_2}$, and $\ket{m_3}$, and its vector
projection onto the $\ket{b_{pqr}} = \ket{b_p} \times \ket{b_q} \times
\ket{b_r}$ subspace.}
\label{fig:diagonal}
\end{figure}

To understand why this is so, let us sum up the $k$ moves that Player
$\sigma$ has made,
\begin{equation}
\ket{m_{\sigma}} = \ket{m_{1\sigma}} + \ket{m_{2\sigma}} + \cdots +
\ket{m_{k\sigma}}
\end{equation}
and check the weights
\begin{equation}
W_{pqr} = |\braket{m_{\sigma} | b_{pqr}} |^2 = |\braket{m_{\sigma} |
b_p }|^2 + |\braket{m_{\sigma}|b_q}|^2 + |\braket{m_{\sigma}|b_r}|^2
\end{equation}
along the eight straight lines on the $3 \times 3$ grid, where $\ket{b_{pqr}} =
\ket{b_p} \times \ket{b_q} \times \ket{b_r}$ is the hypersurface spanned by
$\ket{b_p}$, $\ket{b_q}$, and $\ket{b_r}$.  These can be viewed as the squares
of the scalar projections of the resultant vector $\ket{m_{\sigma}}$ onto the
eight three-dimensional subspaces spanned by $\ket{b_p}$, $\ket{b_q}$, and
$\ket{b_r}$.  Since all quantum moves have to be normalized and orthogonal to
each other, the resultant vector is the diagonal of a $k$-dimensional cube, as
shown in Figure \ref{fig:diagonal}. 

For $k = 3$ moves, the resultant vector $\ket{m_{\sigma}}$ has a length of
$\sqrt{3}$.  Thus, the only way for the square of its scalar projection to be
equal to three is for $\ket{m_{\sigma}}$ to lie entirely within one such
three-dimensional subspace.  It is also impossible for the maximum weight of
three quantum moves to be greater than three.  Since a quantum game offers
infinitely many more moves than the classical game, the set of three successive
moves with resultant vector lying exactly on one of the eight three-dimensional
subspaces is of measure zero.  This explains why no player was found to win
after the third move in our simulations.

\subsection{Opening moves}
\label{sec: OpeningMoves}

To someone learning to play chess formally, the first order of
business is always to learn the various opening moves, and understand
the relative advantages they confer.  An \emph{opening move} is the
first move played in the game.  It is an important move, as it
influences the middle game, and thus also the end game.  In this
subsection, we investigate different opening moves, to better
understand the advantages they confer to Player 1.

For concreteness, let us compare three opening moves: (i) the
\emph{classical opening move}; (ii) the \emph{uniform opening move};
and (iii) the \emph{random opening move}.  In (i), Player 1 always
plays the classical move $\ket{b_5}$ as his first move, whereas in
(ii), Player 1 always start by playing the quantum move
$\frac{1}{\sqrt{9}}\ket{b_1} + \frac{1}{\sqrt{9}}\ket{b_2} + \cdots +
\frac{1}{\sqrt{9}}\ket{b_9}$, which has uniform contribution from all
classical moves.  In (iii), Player 1 plays a random opening move.  For
each opening move, we played 10,000 games for which all subsequent
moves are random quantum moves.  The outcomes are shown in Table
\ref{tab: table2}.

\begin{table}[htbp]
\caption{Outcomes of 10,000 random quantum games each for three
different opening moves: (i) classical; (ii) uniform; and (iii)
random.  Here we show the proportions of wins by Player 1 and Player 2
after move $k$ for both games.  Player 2 has only four moves, so the
number shown for $k = 5$ is the proportion of games ending in a draw.}
\label{tab: table2}
\begin{center}
\begin{tabular}{|c||c|c||c|c||c|c|} \hline
\multirow{3}{*}{Move $k$} & \multicolumn{6}{|c|}{Opening Move} \\
\cline{2-7}
& \multicolumn{2}{|c||}{Classical (\%)} 
& \multicolumn{2}{|c||}{Uniform (\%)} 
& \multicolumn{2}{|c|}{Random (\%)} \\ 
\cline{2-7}
& Player 1 & Player 2 & Player 1 & Player 2 & Player 1 & Player 2 \\ 
\hline \hline
1 & 0 & 0 & 0 & 0 & 0 & 0 \\ \hline
2 & 0 & 0 & 0 & 0 & 0 & 0 \\ \hline
3 & 0 & 0 & 0 & 0 & 0 & 0 \\ \hline
4 & 7.6 & 28.9 & 23.4 & 16.4 & 21.8 & 14.2 \\ \hline
5 & 27.0 & 36.5 & 35.0 & 25.2 & 38.5 & 25.5 \\ \hline
\end{tabular}
\end{center}
\end{table}

As we can see from Table \ref{tab: table2}, the proportions of games won by
Player 1, Player 2, and ending in a tie are very similar for the uniform and
random opening moves, down to the breakdown of proportions of games won after
the fourth and fifth moves.  The situation for the classical opening move,
however, is very different.  While Player 1 still wins more games, Player 2 wins
nearly twice as many games opened with a classical move compared to games opened
with a uniform move or a random move.  This tells us that in the absence of
strategies adopted by Players 1 and 2, a quantum opening move significantly
improves the advantage enjoyed by Player 1.

The geometrical picture behind this quantum advantage is very simple.  The
three-dimensional winning subspace $\ket{b_{pqr}}$ is spanned by the classical
moves $\ket{b_p}$, $\ket{b_q}$, and $\ket{b_r}$.  The moment Player 1 plays the
classical move $\ket{b_p}$, the scalar projection of $\ket{m_1}$ onto
$\ket{b_p}$ saturates at $\braket{b_p|m_1} = 1$.  However, if Player 1 avoids
playing $\ket{b_p}$, the scalar projection $\braket{b_p|m_1}$ can grow with the
number of moves made.  In fact, with an appropriate choice of quantum moves, we
can make $\braket{b_p|m_1} > 1$ after Player 1's second move.  By opening with
$\ket{b_5}$, Player 1 has thus eroded the natural advantage he enjoys in the
game, by limiting the rates at which he is accumulating weights along four of
the eight straight lines.

\subsection{End games}
\label{sect:endgames}

Besides the opening moves, we also learn a game by studying the end
games, whereby the combinatorial complexity of the game is reduced
because there are only a few moves left.  In particular, we studied
end games in which Player 1 is on the verge of winning. To arrive at
an end-game situation, we played random quantum games, and kept those
games where Player 1 wins after his fourth move. We then discard the
moves after Player 1's third move, to obtain an \emph{end game} where
Player 1 has made three moves and Player 2 has made two moves.

Because Player 1 can win on his next move, it is evident that Player 2
must play a \emph{blocking move}.  To stop Player 1 from winning,
Player 2 can play the move Player 1 would use to win, i.e. Player 1's
\emph{winning move}.  Thereafter, Player 1 can no longer play it,
because he is forced to play moves orthogonal to all previous moves.
However, just like in the classical game, Player 1 may have more than
one winning move.  In fact, Player 1 has infinitely many winning moves
within the four-dimensional space of all legal quantum moves
remaining.

Clearly, this manifold of winning moves should be densely distributed
about moves that maximize Player 1's weight along one or more of the
eight straight lines.  To find the \emph{maximizing move} $\ket{x}$
that maximizes Player 1's weight 
\begin{equation}\label{eqn:Wpqr}
W_{pqr} = | \left( \bra{m_1} + \bra{ x } \right) \ket{b_{pqr}} |^2
\end{equation}
along the direction $pqr$, subject to the condition that it orthonormal to all
previous moves, we use the method of Lagrange multipliers.  Here,
$\ket{ m_1 } = \ket{ m_{11} } + \ket{ m_{21} } + \ket{ m_{31} }$ is
the sum of the three moves Player 1 has made.  
Writing out the constraints 
\begin{eqnarray}
\braket{ x|x } = 1, \label{eqn:normalization} \\
\braket{ m_{l\sigma}|x } = 0, \label{eqn:orthogonality}
\end{eqnarray}
explicitly, for $l\sigma = 11, 12, 21, 22, 31$, the simultaneous
equations we need to solve are (see \ref{sec: LagMultMeth} for detail
derivations)
\begin{eqnarray}
-2 \alpha \ket{x} + M \beta + 2 \sum_{s=p,q,r} \ket{b_s}
\left(\braket{b_s | m_1} + \braket{b_s | x } \right) = 0, \label{eqn:max} \\
1 - \braket{ x | x } = 0, \label{eqn:norm} \\
M^T \ket{x} = 0, \label{eqn:orthogonal}
\end{eqnarray}
where $\alpha$ is the Lagrange multiplier for enforcing normalization,
$\beta$ is a $5 \times 1$ vector of Lagrange multipliers for enforcing
orthogonalization, and 
\begin{eqnarray}\label{eqn:PreviousMoves}
M= \left[
\begin{array}{c|c|c|c|c} |m_{11}\rangle & |m_{12}\rangle &
|m_{21}\rangle & |m_{22}\rangle & |m_{31}\rangle \end{array} \right]
\end{eqnarray} 
is a $9 \times 5$ matrix compiling the five previous moves.  Here, $0$
denotes either the scalar, the $5 \times 1$ or the $9 \times 1$ null
vectors depending on the context.

After finding Player 1's eight maximizing moves, and the maximum weights they
are associated with, Player 2 can play the maximizing move with the largest
maximum weight overall as her blocking move.  However, if Player 1 can win along
multiple directions, then Player 2 is sure to lose in the classical tic-tac-toe.
In the quantum tic-tac-toe, Player 2 might be able to take advantage of the
`quantumness' of the game, to simultaneously block all of Player 1's winning
directions.  We evaluated the effectiveness of one such quantum blocking move,
by first sorting the end games according to their \emph{pre-winning weight}.
For end games of a given pre-winning weight $\omega$, we then let Player 2 play
a \emph{weighted blocking move},
\begin{equation}
\ket{y} = \mathcal{N}\left( W_1 \ket{x_1} + W_2 \ket{x_2} + W_3
\ket{x_3} \right),
\end{equation}
consisting of the three best moves $\ket{x_1}$, $\ket{x_2}$, and
$\ket{x_3}$ by Player 1, i.e. the three maximizing moves that gives
the largest \emph{winning weights} $W_1$, $W_2$, and $W_3$.  Here,
$\mathcal{N}$ is a normalization constant we need to compute each time
$\ket{y}$ is constructed, because $\ket{x_1}$, $\ket{x_2}$, and
$\ket{x_3}$ are not necessarily orthogonal to each other.  Finally,
after Player 2 has played $\ket{y}$, we let Player 1 play the
maximizing move $\ket{z_1}$ along the direction $\omega$ is obtained.

\begin{figure}[htbp]
\centering
\subfloat[]{\includegraphics[scale=0.65]{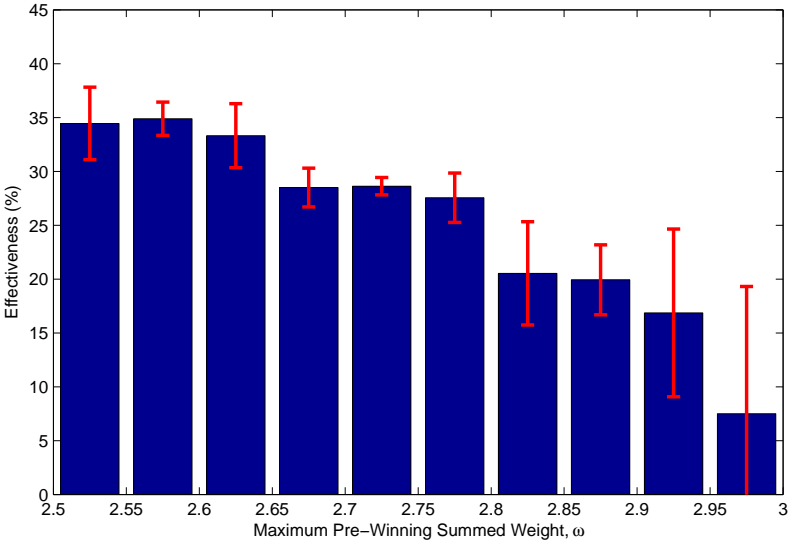} } \\
\subfloat[]{\includegraphics[scale=0.65]{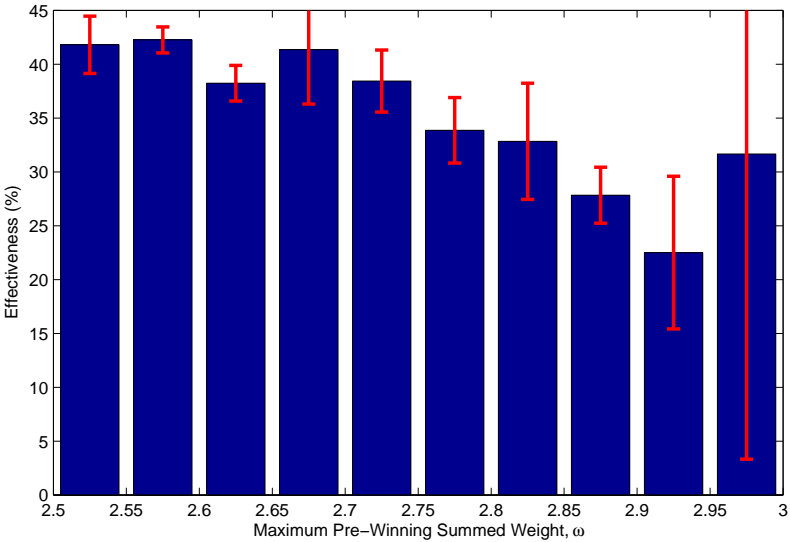} }
\caption{Effectiveness of (a) the random blocking move, and (b) the weighted
blocking move, measured in terms of the proportions of end games successfully
blocked for each pre-winning weight $\omega$.  The weighted blocking move is
about 10\% more effective than the random blocking move.  More importantly, the
weighted blocking move remains highly effective as we approach $\omega = 3$.}
\label{fig: BlockingMove}
\end{figure}

In our simulations, we generated 100,000 end games, and group them
into bins with width $\Delta\omega = 0.05$.  For each bin, we had
Player 2 play the weighted blocking move, as well as a random move not
specifically intended for blocking.  Thereafter, we let Player 1 play
$\ket{z_1}$, before checking whether he has won the game.  As shown in 
Figure \ref{fig: BlockingMove}, we see that the weighted blocking move is
statistically more effective than the random move, not only in terms
of the proportion of end games successfully blocked, but also in terms
of how this proportion falls off as we approach $\omega = 3$.

\section{Deterministic games}
\label{sect:deterministicgames}

After analyzing the end games, we realized that the basic element for playing
the quantum tic-tac-toe is the maximizing move.  We also understood strategic
differences between how Players 1 and 2 were using such a move in the end games.
In essence, Player 2 played a \emph{defensive} third move, seeking only to deny
Player 1 from successfully maximizing his weight.  Following this, Player 1
played an \emph{offensive} fourth move, seeking only to maximize his own weight.
With this insight, we are now able to play the game deterministically, after the
opening move by Player 1.  Our goal is to examine how the outcomes, subject to
different opening moves, depend on following strategies adopted by Players 1 and
2: \begin{enumerate}

\item \emph{Win/block (WB)}.  Player 1 aims to win by playing only offensive
moves, whereas Player 2 plays only blocking moves;

\item \emph{Win-block/block (WBB)}.  Player 1 plays offensive moves, but will
respond with a blocking move if (i) Player 2 will win after the next move,
\emph{and} (ii) he will not win after the present move.  We implement this
blocking condition approximately, by making Player 1 block whenever Player 2's
current pre-winning weight $\omega_2$ exceeds two (and is thus is likely to
exceed three in the next move), and simultaneously his's current pre-winning
weight $\omega_1$ is smaller than $\omega_2$.  Player 2 plays only blocking
moves;

\item \emph{Win/win-block (WWB)}.  Player 1 plays only offensive moves.  Player
2 plays offensive moves, but will respond with a blocking move if (i) Player 1
will win after the next move, \emph{and} (ii) she will not win after the present
move.  Again, we approximate this blocking condition as $\omega_1 > 2$ and
$\omega_1 > \omega_2$ simultaneously;

\item \emph{Win-block/win-block (WBWB)}.  Players 1 and 2 start by playing
offensive moves, but switch over to defensive moves whenever the opponent is on
the verge of winning, and they themselves are not.

\end{enumerate}

To properly define the \emph{offensive move}, let us note that for a given move,
the active player can play eight maximizing moves, one each for directions $pqr
= 123, 456, 789, 147, 258, 369, 159, 357$.  After each of these maximizing moves
are played, the maximum weights that the active player can attain are $W_{123}$,
$W_{456}$, $W_{789}$, $W_{147}$, $W_{258}$, $W_{369}$, $W_{159}$, $W_{357}$
respectively.  The offensive move is the maximizing move associated with the
largest maximum weight overall,
\begin{equation}
W_{\max} = \max\{W_{123},
W_{456}, W_{789}, W_{147}, W_{258}, W_{369}, W_{159}, W_{357}\}.
\end{equation}
As defined in the previous section, the \emph{defensive move} is the
weighted superposition of the opponent's three best maximizing moves.

Because of the normalization constraint, we have to solve a nonlinear system of
simultaneous equations to find each maximizing move.  This is done numerically
using a nonlinear optimization routine in MATLAB, using random initial guesses.
Depending on our initial guess, we can converge to a global maximizing move, or
to stationary solutions that do not maximize the active player's weight along
the given direction.  Therefore, for each direction, we solve for stationary
moves starting with 20 initial guesses.  We then select the stationary move with
the maximum weight, and perform a second-derivative test on it.  If it is
locally maximum, we accept the stationary move as our maximizing move.  Although
this procedure is not guaranteed to always find the globally maximizing move, we
find it giving reliable results in practice.  Details on the second-derivative
test can be found in \ref{sect:Hessian}.

Before we move on to discuss our results, we would like to remark that
though the strategies are deterministic, the games do not progress
deterministically, because of the random initial guesses used to solve
for maximizing moves.  This probabilitistic progress of the games is
most prominent for highly degenerate games, like those opened with a
classical move or a uniform move.  Play-by-play analysis of the
deterministic quantum games for different strategies can be found at
Ref.~\cite{DigiRep}.  In this paper, we will focus on generic
outcomes shown in Table \ref{tab: table3} for the different
strategies, subject to different opening moves.

\begin{table}[htbp]
\begin{center}
\caption{Outcomes of deterministic quantum games each for the
Win/Block (WB), Win-Block/Block (WBB), Win/Win-Block (WWB), and
Win-Block/Win-Block (WBWB) strategies, subject to the classical,
uniform, and random opening moves.  The move number is not listed, but
increases from $k = 1$ to $k = 5$ downwards.  Player 2 has only four
moves, so the proportional shown in the fifth row under Player 2 is
the proportion of games that ended in a draw.  Also, not all 10,000
games were played to completion for each strategy pair and opening
move, because the active player fails to find maximizing moves at some
point in the game.  The number at the last row of each strategy pair
indicates how many games ended prematurely because of this problem.
The proportions shown in the table are computed from the successfully
completed games.}
\label{tab: table3}
\begin{tabular}{|c||c|c||c|c||c|c|} \hline
\multirow{3}{*}{Strategy} & \multicolumn{6}{|c|}{Opening Move} \\
\cline{2-7}
& \multicolumn{2}{|c||}{Classical (\%)} 
& \multicolumn{2}{|c||}{Uniform (\%)} 
& \multicolumn{2}{|c|}{Random (\%)} \\ 
\cline{2-7}
& Player 1 & Player 2 & Player 1 & Player 2 & Player 1 & Player 2 \\ \hline \hline
\multirow{6}{*}{WB} & 0 &  0 &  0 &  0 &  0 &  0 \\ \cline{2-7}
& 0 &  0 &  0 &  0 &  0 &  0 \\ \cline{2-7}
& 0 &  0 &  0 &  0 &  0 &  0 \\ \cline{2-7}
& 0 &  22.4 &  68.4 &  2.9 &  40.5 &  6.2 \\ \cline{2-7}
& 5.1 &  72.5 &  6.2 &  22.5 &  5.0 &  48.3 \\ \cline{2-7}
& \multicolumn{2}{|c||}{183 games} & \multicolumn{2}{|c||}{2186 games} & \multicolumn{2}{|c|}{5279 games} \\ \hline
\multirow{6}{*}{WBB} & 0 &     0 &     0 &     0 &     0 &     0 \\ \cline{2-7} 
& 0 &     0 &     0 &     0 &     0 &     0 \\ \cline{2-7} 
& 0 &     0 &     0 &     0 &     0 &     0 \\ \cline{2-7} 
& 0 &    21.2 &    68.1 &     3.1 &    21.5 &    15.0 \\ \cline{2-7} 
& 6.3 &    72.5 &     6.1 &    22.6 &    10.6 &    52.9 \\ \cline{2-7}
& \multicolumn{2}{|c||}{176 games} & \multicolumn{2}{|c||}{2139 games} & \multicolumn{2}{|c|}{5543 games} \\ \hline
\multirow{6}{*}{WWB} & 0.0 & 0.0 & 0.0 & 0.0 & 0.0 & 0.0 \\ \cline{2-7} 
& 0.0 &   0.0 &   0.0 &   0.0 &   0.0 &   0.0 \\ \cline{2-7} 
& 13.3 &   0.5 &   0.0 &   0.0 &   0.0 &   0.0 \\ \cline{2-7} 
& 0.9 &  15.5 &  44.6 &   1.5 &  51.4 &   6.0 \\ \cline{2-7} 
& 4.8 &  65.0 &  2.8 &  51.2 &  3.6 &  39.0 \\ \cline{2-7}
& \multicolumn{2}{|c||}{420 games} & \multicolumn{2}{|c||}{1605 games} & \multicolumn{2}{|c|}{2083 games} \\ \hline
\multirow{6}{*}{WBWB} & 0.0 & 0.0 & 0.0 & 0.0 &  0.0 &  0.0 \\ \cline{2-7} 
& 0.0 &  0.0 &  0.0 &  0.0 &  0.0 &  0.0 \\ \cline{2-7} 
& 0.0 &  0.0 &  0.0 &  0.0 &  0.0 &  0.0 \\ \cline{2-7} 
& 2.0 &  24.8 &  15.7 &  6.7 &  43.7 &  6.3 \\ \cline{2-7} 
& 10.9 &  62.3 &  41.5 &  36.1 &  12.8 &  37.1 \\ \cline{2-7}
& \multicolumn{2}{|c||}{442 games} & \multicolumn{2}{|c||}{1575 games} & \multicolumn{2}{|c|}{3719 games} \\ \hline
\end{tabular}
\end{center}
\end{table}

\subsection{Comparison against the deterministic classical game}

From Table \ref{tab: table3}, we see that the deterministic quantum
tic-tac-toe do not always end up in a draw, even for the classical
opening move, when the proportions of games ending in a draw is
highest (around 70\%), whatever the strategy pair.  This is a clear
departure from the classical tic-tac-toe, where all deterministic
games must end in a draw \cite{Berlekamp2003WinningWays}.  Between the
two quantum opening moves, the proportion of tied games is generally
lower for games opened with the uniform move than for games opened
with the random move.  We expect this, because the uniform opening
move confers the maximum quantum advantage on Player 1, who would go
on to win most of these deterministic games.

What is perhaps more surprising, is Player 2 winning more deterministic games
than Player 1, whatever the strategy pair, when these games are opened with the
classical move!  We know of no classical two-player games whereby Player 2 owns
the advantage.  It turns out that the reason Player 1 does poorly, after opening
with the classical move, is the same for deterministic games as it is for random
games.  After saturating the scalar projection $\braket{m_1 | b_5}$ with the
opening move, Player 1 effectively traded away his ability to more rapidly
increase his weights along four out of eight directions with further moves.
This loss of advantage by Player 1 is extremely pronounced in the WB and WBB
games, from winning over 30\% of random quantum games opened with the classical
move $\ket{b_5}$, to about 5\% in deterministic games opened with $\ket{b_5}$.
Since Player 2 is playing defensively in these two class of games, her winning
proportions did not increase over that in the random games.  The sharp drop in
Player 1's winning proportions is thus a testimony on how effective the quantum
blocking move is.

\subsection{Comparison between opening moves}

In contrast to the classical opening move, the uniform and random opening moves
confer immense advantage onto Player 1, when we compare their outcomes against
those of random quantum games opened with the same moves.  Player 2 went from
winning about 15\% of the random games to winning about 3--6\% in the
deterministic games.  The only exception is WBB games opened with a random move,
where Player 2 apparently suffers no further quantum disadvantage.  Comparing
Tables \ref{tab: table2} and \ref{tab: table3}, we find Player 1 wins more of
his random games after $k = 5$ moves, but most of his WB, WBB, WWB games after
$k = 4$ moves.  This shows that the quantum opening move is an effective move
for Player 1, when playing strategically.

We were also surprised to find Player 1 winning 13.3\% of the WWB games opened
with the classical move after the third move.  Upon checking the games play by
play for this strategy pair, we found that the pre-winning weight of Player 1
should always be $\omega_1 = 2$.  Depending on numerical truncation errors, the
numerical value of $\omega_1$ either just fails or just succeeds to trigger the
criteria for Player 2 to start blocking.  In the former, Player 2 plays an
offensive second move, leaving Player 1 unhampered to play a winning third move.
In the latter, Player 2 plays a blocking second move, effectively denying Player
1 of his third-move win.  Because of the integer nature of the classical opening
move, the numerical truncation errors associated with $\omega_1$ is smaller than
those associated with $\omega_2$, after the same number of moves.  Thus, Player
1's third move in WBWB games opened with the classical move is almost always a
blocking move.  This explains why Player 1 is not observed to win after three
moves in such games.

\subsection{Comparison between different strategies}

With the classical opening move, Player 1 seriously disadvantaged
himself.  His winning proportion is lowest when he plays to win, while
Player 2 plays to block.  We might be tempted to think that this is
because he fails to block Player 2 when she is on the verge of
winning.  But when Player 1 plays to win, but also block Player 2
whenever necessary, his winning proportion increases only slightly,
from 5.1\% to 6.3\%.  In contrast, when Player 2 decides to start with
an offensive move, and block only when necessary, in the WWB and WBWB
games, Player 1 is no longer quite as disadvantaged.  This tells us
that the major factor affecting Player 1's fortune is whether Player 2
choose to start defensively or offensively.

This same pattern is repeated for the quantum opening moves.  Player 1
does no worse, or slightly better when he also blocks, than when he
single-mindedly plays to win, for the same Player 2 strategy.  On the
other hand, Player 2 is worse off if she also plays to win, than when
she single-mindedly blocks, if she is playing against a purely
offensive Player 1.  She fares better with mixed offensive-defensive
moves, than with purely defensive moves, however, if Player 1 also
plays mixed offensive-defensive moves.

\section{Conclusions}
\label{sect:conclusions}

To conclude, we have in this paper introduced a minimalistic quantization of the
classical tic-tac-toe, by admitting quantum moves which are arbitrary
superpositions of the classical moves.  We require our quantum moves to be
orthonormal to all previous moves, and also for the sum of squares of resultant
amplitudes to exceed three along any straight line of three cells for a player
to win, so that our quantum tic-tac-toe reduces properly to the classical
tic-tac-toe.  Playing the quantum game first randomly and then
deterministically, we find that unlike the classical game, the deterministic
quantum tic-tac-toe does not always end in a draw.  Furthermore, unlike most
classical two-player games of no chance, both players can win in the
deterministic quantum game.  More interestingly, in both random and
deterministic quantum games, we see that Player 1 enjoys an overwhelming quantum
advantage when he opens with a quantum move.  This advantage, which is lost when
Player 1 opens with a classical move, has a very simple geometrical
interpretation in terms of the projection of the resultant move onto the
classical winning subspaces.  Finally, the biggest contrast between the
classical and quantum tic-tac-toes must surely be the effective quantum blocking
move that the defending player can play.  In fact, a defensive strategy based
solely on such a quantum blocking move is the strategy of choice for Player 2,
for most strategies that Player 1 adopts.

While the quantum tic-tac-toe does not properly describe the quantum information
transfer scenario developed in the Introduction, we believe some generic results
obtained for the former should also apply in the latter.  For instance, we
believe Alice will also enjoy a huge quantum advantage with a uniform opening
move, if we imagine she has multiple quantum channels through which she can
transmit to Bob.  This move is the least informative, and Eve would have to
guess which quantum channels will ultimately be used to transmit the message to
Bob, in order to come up with a blocking move.  Certainly, Alice should not
first attempt to transmit a classical bit utilizing just one channel, because
she will almost certainly lose the advantage she naturally enjoys as Player 1.
Eve can learn something from this paper as well.  If the transmissions by Alice
as to be understood as purely offensive moves, Eve should adopt a pure quantum
jamming strategy by playing quantum blocking moves.  She should not succumb to
the temptation to also intercept the message, which we can interpret as an
offensive move, because she is not likely to do any better with such a mixed
strategy.

Like the quantum information transfer scenario, the multiplayer portfolio
optimization game idealized in the Introduction differs from the quantum
tic-tac-toe in many important aspects.  In particular, both the multiplayer
portfolio optimization game and the quantum information transfer game are not
subjected to stringent orthonormality constraints.  Nevertheless, we believe the
generic lessons learnt from the quantum tic-tac-toe will apply even in this
significantly more complex quantum game.  To prevent competitors from concerted
or inadvertent sabotage, a fund manager should play a uniform move by maximally
diversifying his portfolio.  This is because adjustments to such a portfolio
yields the least information for other fund managers to act upon, and therefore
its optimality is least susceptible to malicious attacks.  Should a fund manager
suspect intentional attacks to his portfolio by multiple players, we also expect
the quantum blocking move to be highly effective.  We believe such a `defensive'
strategy will help a fund fare better during a financial crisis, where the
cascading loss-cutting measures adopted by other funds can be seen as a
coordinated assault on its position.

Finally, we note that in the duel between grandmasters, there is the
additional element of timing in the strategic game play.  For example,
an effective move can be planted ahead of time, and its effectiveness
enhanced by subsequent moves.  Another example would be, at times
where a defensive move seems inevitable, a grandmaster can force his
opponent's hand by playing an offensive move elsewhere, and then
return leisurely to play the defensive move.  In our quantum
tic-tac-toe, the game complexity is not high enough for such
situations to arise.  A future topic of research would be to quantize
a more complex two-player game, where these timing situations do
arise, and then explore game-theoretically how different the
outcome might be if quantum moves are made available.

\ack This work is supported by startup grant SUG 19/07 provided by the
Nanyang Technological University.  We thank Lock Yue Chew, Pinaki
Sengupta, and Yon Shin Teo for discussions.

\appendix
\section{Method of Lagrange multipliers}
\label{sec: LagMultMeth}
\setcounter{section}{1}

In Section \ref{sect:endgames}, the method of Lagrange multipliers was used to
find the maximizing move $\ket{x}$ along a given direction $pqr$.  In this
appendix, we will describe how we obtain the simultaneous equations
(\ref{eqn:max}), (\ref{eqn:norm}), and (\ref{eqn:orthogonal}).  In the method of
Lagrange multipliers, if $f(x, y)$ is the function we wish to maximize, subject
to the constraints, $g(x,y)=c$ and $h(x,y)=d$, we introduce the \emph{Lagrange
function},
\begin{equation}\label{eq: LagFunc}
\Lambda(x,y,\alpha,\beta) = f(x,y) + \alpha (g(x,y)-c) + \beta (h(x,y)-d) 
\end{equation}
where $\alpha$ and $ \beta $ are the \emph{Lagrange multipliers}.  To maximize
$\Lambda(x, y, \alpha, \beta)$, we partial differentiate $\Lambda(x, y, \alpha,
\beta)$ with respect to $x$ and $y$, as well as $\alpha$ and $\beta$, and set
the partial derivatives $\pa{\Lambda}{x}$, $\pa{\Lambda}{y}$,
$\pa{\Lambda}{\alpha}$, $\pa{\Lambda}{\beta}$ to zero.

In the end-game situation discussed in Section \ref{sect:endgames}, Player 1 has
made his third move, and we would like to maximize his weight along the
direction $pqr$, using a normalized move orthogonal to all previous moves.  In
this situation, the function we would like to maximize is the weight $W_{pqr}$,
given in Eqn.~(\ref{eqn:Wpqr}), and the normalization and orthogonality
constraints are given by Eqn.~(\ref{eqn:normalization}) and
Eqn.~(\ref{eqn:orthogonality}) respectively.  Using Eqn.~(\ref{eq: weight}), we
can write the weight of Player 1 along $pqr$ after the maximizing move
explicitly as
\begin{equation}
W_{pqr}^{41} = \sum_{i=p,q,r} \left(\sum_{l=1}^3 v_{il1} + x_i \right)^2.
\end{equation}
We can also write the normalization and orthogonality constraints out explicitly
as
\begin{equation}
\sum^{9}_{i=1} x_i^2 = 1, \quad
\sum_{i} v_{il\sigma}x_{i} = 0,
\end{equation}
where $x_{i}$ is the $i$th amplitude of $|x\rangle$, and $l\sigma = 11, 12, 21,
22, 31$.  With these, our Lagrange function becomes 
\begin{equation}\label{eqn:LagrangeFunction}
\Lambda = \sum_{i=p,q,r} \left(\sum_{l=1}^3 v_{il\sigma} + x_i \right)^2 +
\alpha \left (1-\sum^{9}_{i=1} x_i^2\right ) + \sum_{\{l\sigma\}}
\beta_{l\sigma} \left (\sum_{i} v_{il\sigma}x_{i}\right ),
\end{equation}
using a total of six Lagrange multipliers, $\alpha$ to enforce normalization,
and five $\beta_{l\sigma}$ to enforce orthogonality with respect to each of the
five previous moves.

Differentiating the Lagrange function with respect to $x_i$, we find
\begin{equation}
\frac{\partial \Lambda}{\partial x_i} = - 2 \alpha x_i + \sum_{\{l\sigma\}}
\beta_{l\sigma} v_{il\sigma} = 0
\end{equation}
if $i \neq p,q,r$.  If $i$ is $p$, $q$, or $r$, then $\pa{\Lambda}{x_i}$ has an
extra term $2 \left(\sum_{l=1}^3 v_{il\sigma} + x_i \right) $ arising from the
first term in Eqn.~(\ref{eq: LagFunc}).  We combine these two types of partial
derivatives by writing
\begin{equation}\label{eq: dLagFunc}
\frac{\partial \Lambda}{\partial x_i} = -2\alpha x_i + \sum_{\{l\sigma\}}
\beta_{l\sigma} v_{il\sigma} + 2\left(\sum_{l=1}^3 v_{il\sigma} + x_i
\right)_{pqr} = 0
\end{equation}
where the subscript $pqr$ in the last term indicates that we only add the last
term if $i=p,q$ or $r$.  This becomes Eqn.~(\ref{eqn:max}) when written in
matrix-vector form.  Eqn.~(\ref{eqn:norm}) and Eqn.~(\ref{eqn:orthogonal}) are
simply $\pa{\Lambda}{\alpha} = 0$, the normalization constraint, and
$\pa{\Lambda}{\beta_{l\sigma}} = 0$, the orthogonality constraints, written in
matrix-vector form.

\section{Hessian matrix and second-derivative test}
\label{sect:Hessian}

To do the second-derivative test for the maximizing move, we first evaluate the
Hessian matrix
\begin{equation}
\renewcommand{\arraystretch}{1.5}
\fl
H(\Lambda)=
\begin{bmatrix}
\frac{\partial^2 \Lambda}{\partial x_1^2} & 
\frac{\partial^2 \Lambda}{\partial x_1\,\partial x_2} & \cdots & 
\frac{\partial^2 \Lambda}{\partial x_1\,\partial x_9} &
\Pab{\Lambda}{x_1}{\alpha} & \Pab{\Lambda}{x_1}{\beta_{11}} &
\Pab{\Lambda}{x_1}{\beta_{12}} & \dots & \Pab{\Lambda}{x_1}{\beta_{l\sigma}} \\
\frac{\partial^2 \Lambda}{\partial x_2\,\partial x_1} & 
\frac{\partial^2 \Lambda}{\partial x_2^2} & \cdots & 
\frac{\partial^2 \Lambda}{\partial x_2\,\partial x_9} &
\Pab{\Lambda}{x_2}{\alpha} &
\Pab{\Lambda}{x_2}{\beta_{11}} &
\Pab{\Lambda}{x_2}{\beta_{12}} & \cdots &
\Pab{\Lambda}{x_2}{\beta_{l\sigma}} \\
\vdots & \vdots & \ddots & \vdots & \vdots & \vdots & \vdots & \ddots & \vdots \\
\frac{\partial^2 \Lambda}{\partial x_9\,\partial x_1} & 
\frac{\partial^2 \Lambda}{\partial x_9\,\partial x_2} & \cdots & 
\frac{\partial^2 \Lambda}{\partial x_9^2} &
\Pab{\Lambda}{x_9}{\alpha} &
\Pab{\Lambda}{x_9}{\beta_{11}} &
\Pab{\Lambda}{x_9}{\beta_{12}} & \cdots &
\Pab{\Lambda}{x_9}{\beta_{l\sigma}} \\
\Pab{\Lambda}{\alpha}{x_1} & 
\Pab{\Lambda}{\alpha}{x_2} & \cdots &
\Pab{\Lambda}{\alpha}{x_9} & 
\Paa{\Lambda}{\alpha} &
\Pab{\Lambda}{\alpha}{{\beta_{11}}} &
\Pab{\Lambda}{\alpha}{{\beta_{12}}} & \cdots &
\Pab{\Lambda}{\alpha}{{\beta_{l\sigma}}} \\
\Pab{\Lambda}{\beta_{11}}{x_1} &
\Pab{\Lambda}{\beta_{11}}{x_2} & \cdots &
\Pab{\Lambda}{\beta_{11}}{x_9} &
\Pab{\Lambda}{\beta_{11}}{\alpha} &
\Paa{\Lambda}{\beta_{11}} &
\Pab{\Lambda}{\beta_{11}}{\beta_{12}} & \cdots &
\Pab{\Lambda}{\beta_{11}}{\beta_{l\sigma}} \\
\Pab{\Lambda}{\beta_{12}}{x_1} &
\Pab{\Lambda}{\beta_{12}}{x_2} & \cdots &
\Pab{\Lambda}{\beta_{12}}{x_9} &
\Pab{\Lambda}{\beta_{12}}{\alpha} &
\Pab{\Lambda}{\beta_{12}}{\beta_{11}} &
\Paa{\Lambda}{\beta_{12}} & \cdots &
\Pab{\Lambda}{\beta_{12}}{\beta_{l\sigma}} \\
\vdots & \vdots & \ddots & \vdots & \vdots & \vdots & \vdots & \ddots & \vdots \\
\Pab{\Lambda}{\beta_{l\sigma}}{x_1} &
\Pab{\Lambda}{\beta_{l\sigma}}{x_2} & \cdots &
\Pab{\Lambda}{\beta_{l\sigma}}{x_9} &
\Pab{\Lambda}{\beta_{l\sigma}}{\alpha} &
\Pab{\Lambda}{\beta_{l\sigma}}{\beta_{11}} &
\Pab{\Lambda}{\beta_{l\sigma}}{\beta_{12}} & \cdots &
\Paa{\Lambda}{\beta_{l\sigma}}
\end{bmatrix}
\end{equation}
of the Lagrange function given in Eqn.~(\ref{eqn:LagrangeFunction}).

Since the Lagrange function $\Lambda(x_1, x_2, \dots, x_9, \alpha, \beta_{11},
\beta_{12}, \dots, \beta_{l\sigma})$ does not contain cross terms of the form
$x_i x_j$, the $9 \times 9$ submatrix in $H(\Lambda)$ is diagonal, with diagonal
matrix elements
\begin{equation}
H_{ii}(\Lambda) = \frac{\partial^2 \Lambda}{\partial x_i^2} = -2\alpha +\left( 2
\right)_{pqr}.
\end{equation}
Differentiating Eqn.~(\ref{eq: dLagFunc}) with respect to $\alpha$ and $\beta$,
we will also get
\begin{eqnarray}
\frac{\partial^2 \Lambda}{\partial \alpha \partial x_i} &= -2 x_i, \\
\frac{\partial^2 \Lambda}{\partial \beta_{l\sigma} \partial x_i} &= v_{il\sigma}
\end{eqnarray}
respectively.  Finally, we see that there are neither quadratic or cross terms
involving $\alpha$ and $\beta$ in the Lagrange function,
Eqn.~(\ref{eqn:LagrangeFunction}), and thus the second partial derivatives of
$\Lambda(x_1, x_2, \dots, x_9, \alpha, \beta_{11}, \beta_{12}, \dots,
\beta_{l\sigma})$ with respect to the Lagrange multipliers are always zero.  The
Hessian matrix is thus
\begin{equation}
H(\Lambda) = 
\begin{bmatrix}
A & -2|x\rangle & M \\
-2|x\rangle^T & \multicolumn{2}{c}{\multirow{2}{*}{$\mathcal{O}$}}\\
M^T
\end{bmatrix},
\end{equation}
where $A$ is a $9 \times 9$ diagonal matrix, with all the diagonal entries being
$-2\alpha$, except the $p$th, $q$th and $r$th diagonal entries, which are
$-2\alpha + 2$.  The matrix $M$ is the matrix compiling all previous moves
defined in Eqn.~(\ref{eqn:PreviousMoves}), while $\mathcal{O}$ is a $(k+1)
\times (k+1)$ null matrix, $k$ being the total number of moves made by both
players. 

We then evaluate the Hessian matrix $H(\Lambda)$ at the optimal values $(x_1^*,
x_2^*, \dots, x_9^*;$ $\alpha^*$, $\beta_{11}^*, \beta_{12}^*, \dots,
\beta_{l\sigma}^*)$ of the maximizing move, before diagonalizing it to check if
the maximizing move does indeed maximize the weight of the active player.  In
unconstrained optimization within a $d$-dimensional space of parameters, we must
have $d$ negative eigenvalues, for the given optimal point to be locally
maximum.  In constrained optimization, each constraint defines a hypersurface.
The constrained optimal point need not be locally maximum along directions
normal to these constraint hypersurfaces, since we are not allowed to venture
off these hypersurfaces anyway.  If $k$ moves have already been played, there
will be $k$ normal directions.  The eigenvalues of $H(\Lambda)$ associated with
eigenvectors lying within the space spanned by these $k$ normal vectors need not
be negative.  Hence, a maximizing move is locally maximum if $H(\Lambda)$ has at
least $n = 9 - k$ negative eigenvalues, where $n$ is the number of moves
remaining.  Only deterministic quantum games for which all moves are locally
maximizing are reported in this paper (see Table \ref{tab: table3}).


\section*{References}

\begin{thebibliography}{99}

\bibitem{Bouwmeester1997Nature390p575} D. Bouwmeester, J.-W. Pan, K.
Mattle, M. Eibl, H. Weinfurter, and A. Zeilinger, Experimental quantum
teleportation, Nature \textbf{390}, 575--579 (1997).

\bibitem{Bennett1993PhysRevLett70p1895} C. H. Bennett, G. Brassard, C.
Cr\'epeau, R. Jozsa, A. Peres, and W. K. Wootters, Teleporting an
unknown quantum state via dual classical and Einstein-Podolsky-Rosen
channels, Physical Review Letters \textbf{70(13)}, 1895--1899 (1993).

\bibitem{Nielsen1998Nature396p52} M. A. Nielsen, E. Knill, and R.
Laflamme, Complete quantum teleportation using nuclear magnetic
resonance, Nature \textbf{396}, 52--55 (1998).

\bibitem{Furusawa1998Science282p706} A. Furusawa, J. L. S\o{}rensen,
S. L. Braunstein, C. A. Fuchs, H. J. Kimble, and E. S. Polzik,
Unconditional quantum teleportation, Science \textbf{282}, 706--709
(1998).

\bibitem{Pan2001Nature410p1067} J.-W. Pan, C. Simon, \v{C}. Brukner,
and A. Zeilinger, Entanglement purification for quantum communication,
Nature \textbf{410}, 1067--1070 (2001).

\bibitem{Riebe2004Nature429p734} M. Riebe, H. H\"affner, C. F. Roos,
W. H\"ansel, J. Benhelm, G. P. T. Lancaster, T. W. K\"orber, C.
Becher, F. Schmidt-Kaler, D. F. V. James, and R. Blatt, Deterministic
quantum teleportation with atoms, Nature \textbf{429}, 734--737
(2004).

\bibitem{Barrett2004Nature429p737} M. D. Barrett, J. Chiaverini, T.
Schaetz, J. Britton, W. M. Itano, J. D. Jost, E. Knill, C. Langer, D.
Leibfried, R. Ozeri, and D. J. Wineland, Deterministic quantum
teleportation of atomic qubits, Nature \textbf{429}, 737--739 (2004).

\bibitem{Chaneliere2005Nature438p833} T. Chaneli\`ere, D. N.
Matsukevich, S. D. Jenkins, S.-Y. Yan, T. A. B. Kennedy, and A.
Kuzmich, Storage and retrieval of single photons transmitted between
remote quantum memories, Nature \textbf{438}, 833--836 (2005).

\bibitem{Bennett2000Nature404p247} C. H. Bennett and D. P. DiVincenzo,
Quantum information and computation, Nature \textbf{404}, 247--255
(2000).

\bibitem{Galindo2002RevModPhys74p347} A. Galindo and M. A.
Mart\'in-Delgado, Information and computation: Classical and quantum
aspects, Reviews of Modern Physics \textbf{74}, 347--423 (2002).

\bibitem{Braunstein2005RevModPhys77p513} S. L. Braunstein and P. van
Loock, Quantum information with continuous variables, Reviews of
Modern Physics \textbf{77}, 513--577 (2005).

\bibitem{Bennett1984IntConfCompSysSigProc} C. H. Bennett and G.
Brassard, Quantum cryptography: Public key distribution and coin
tossing, Proceedings of the IEEE International Conference on
Computers, Systems \& Signal Processing (Bangalore, India, Dec 10--12,
1984), 1984.

\bibitem{Ekert1991PhysRevLett67p661} A. K. Ekert, Quantum cryptography
based on Bell's theorem,  Physical Review Letters \textbf{67(6)},
661--663 (1991).

\bibitem{Bennett1992PhysRevLett68p557} C. H. Bennett, G. Brassard, and
N. D. Mermin, Quantum cryptography without Bell's theorem, Physical
Review Letters \textbf{68(5)}, 557--559 (1992).

\bibitem{Bennett1992PhysRevLett68p3121} C. H. Bennett, Quantum
cryptography using any two nonorthogonal states, Physical Review
Letters \textbf{68(21)}, 3121--3124 (1992).

\bibitem{Gisin2002RevModPhys74p145} N. Gisin, G. Ribordy, W. Tittel,
and H. Zbinden, Quantum cryptography, Reviews of Modern Physics
\textbf{74(1)}, 145--195 (2002).

\bibitem{Meyer1999PhysRevLett82p1052} D. A. Meyer, Quantum strategies,
Physical Review Letters \textbf{82(5)}, 1052--1055 (1999).

\bibitem{Eisert1999PhysRevLett83p3077} J. Eisert, M. Wilkens, and M.
Lewenstein, Quantum games and quantum strategies, Physical Review
Letters

\bibitem{Du2002ChinPhysLett19p1221} J.-F. Du, H. Li, X.-D. Xu, X.-Y.
Zhou, R.-D. Han, Multi-player and multi-choice quantum game, Chinese
Physics Letters \textbf{19(9)}, 1221--1224 (2002).

\bibitem{Zhao2004ChinPhysLett21p1421} H.-J. Zhao and X.-M. Fang,  Does
the quantum player always win the classical one? Chinese Physics
Letters \textbf{21(8)}, 1421--1424 (2004).

\bibitem{Aharon2008PhysRevA77e052310} N. Aharon and L. Vaidman,
Quantum advantages in classically defined tasks, Physical Review A
\textbf{77(5)}, 052310 (2008).

\bibitem{Johnson2001PhysRevA63e020302R} N. F. Johnson, Playing a
quantum game with a corrupted source, Physical Review A
\textbf{63(2)}, 020302(R) (2001).

\bibitem{Chen2002PhysRevA65e052320} J.-L. Chen, L. C. Kwek, and C. H.
Oh, Noisy quantum game, Physical Review A \textbf{65(5)}, 052320
(2002).

\bibitem{Guinea2003JPhysAMathGen36pL197} F. Guinea and M. A.
Mart\'in-Delgado, Quantum Chinos game: winning strategies through
quantum fluctuations, Journal of Physics A: Mathematical and General
\textbf{36(13)}, L197--L204 (2003).

\bibitem{Flitney2005JPhysAMathGen38p449} A. P. Flitney and D. Abbott,
Quantum games with decoherence, Journal of Physics A: Mathematical and
General \textbf{38(2)}, 449--460 (2005).

\bibitem{Du2005JPhysAMathGen38p1559} J. Du, C. Ju, and H. Li, Quantum
entanglement helps in improving economic efficiency, Journal of
Physics A: Mathematical and General \textbf{38(7)}, 1559--1565 (2005).

\bibitem{Yukalov2010EurPhysJB71p533} V. I. Yukalov and D. Sornette,
Physics of risk and uncertainty in quantum decision making, The
European Physical Journal B \textbf{71(4)}, 533--548 (2009).

\bibitem{Yukalov2010TheoryDecision} V. I. Yukalov and D. Sornette,
Decision theory with prospect interference and entanglement, Theory
and Decision, Feb 2010.  DOI:10.1007/s11238-010-9202-y.

\bibitem{Iqbal2002PhysRevA65e022306} A. Iqbal and A. H. Toor, Quantum
mechanics gives stability to a Nash equilibrium, Physical Review A
\textbf{65(2)}, 022306 (2002).

\bibitem{Lee2003PhysRevA67e022311} C. F. Lee and N. F. Johnson,
Efficiency and formalism of quantum games, Physical Review A
\textbf{67(2)}, 022311 (2003).

\bibitem{Nawaz2004JPhysAMathGen37p11457} A. Nawaz and A. H. Toor,
Generalized quantization scheme for two-person non-zero sum games,
Journal of Physics A: Mathematical and General \textbf{37(47)},
11457--11464 (2004).

\bibitem{Arfi2005TheoryDecision59p127} B. Arfi, Resolving the Trust
Predicament: A Quantum Game-theoretic Approach, Theory and Decision
\textbf{59(2)}, 127--174 (2005).

\bibitem{Ozdemir2007NewJPhys9p43} S. K. \"Ozdemir, J. Shimamura, and
N. Imoto, A necessary and sufficient condition to play games in
quantum mechanical settings, New Journal of Physics \textbf{9}, 43
(2007).

\bibitem{Ichikawa2008JPhysAMathTheor41p135303} T. Ichikawa, I.
Tsutsui, and T. Cheon, Quantum game theory based on the Schmidt
decomposition, Journal of Physics A: Mathematical and Theoretical
\textbf{41(13)}, 135303 (2008).

\bibitem{Berlekamp2003WinningWays} E. R. Berlekamp, J. H. Conway, and
R. K. Guy, \textsl{Winning Ways for Your Mathematical Plays, volume
3, 2nd edition}, A K Peters (Massachussets, USA), 2003.

\bibitem{Goff2002Proc38thAIAAConf} A. Goff, D. Lehmann, and J. Siegel,
Quantum tic-tac-toe, spooky-coins \& magic-envelopes, as metaphors for
relativistic quantum physics, Proceedings of the 38th
AIAA/ASME/SAE/ASEE Joint Propulsion Conference and Exhibit
(Indianapolis, USA, 7-10 July 2002), 2002.

\bibitem{Goff2006AmJPhys74p962} A. Goff, Quantum tic-tac-toe: A
teaching metaphor for superposition in quantum mechanics, American
Journal of Physics \textbf{74(16)}, 962--973 (2006).

\bibitem{DigiRep} J. N. Leaw, \emph{Quantum Tic Tac Toe}, final year project
thesis, School of Physical and Mathematical Sciences, Nanyang Technological
University.  Available at URL: \url{http://hdl.handle.net/10356/40799}.

\end{thebibliography}
\end{document}